%% file: ms.tex
\documentclass[onecolumn]{emulateapj}

\begin{document}

\title{Optimal Coaddition of Imaging Data for Rapidly Fading Gamma-Ray
  Burst Afterglows}
\author{
A.\ N.\ Morgan\altaffilmark{1,2},
D.\ E.\ Vanden~Berk\altaffilmark{1},
P.\ W.\ A.\ Roming\altaffilmark{1},
J.\ A.\ Nousek\altaffilmark{1},
T.\ S.\ Koch\altaffilmark{1},
A.\ A.\ Breeveld\altaffilmark{3},
M.\ de Pasquale\altaffilmark{3},
S.\ T.\ Holland\altaffilmark{4,5},
N.\ P.\ M.\ Kuin\altaffilmark{3},
M.\ J.\ Page\altaffilmark{3},
M.\ Still\altaffilmark{3}
}


\altaffiltext{1}{Department of Astronomy and Astrophysics,
  Pennsylvania State University, University Park, PA, USA.}
\altaffiltext{2}{Institute of Astronomy, University of Cambridge, Cambridge,
  UK.}
\altaffiltext{3}{Mullard Space Science Laboratory, Dorking, Surrey, UK.}
\altaffiltext{4}{NASA Goddard Space Flight Center, Greenbelt, MD, USA.}
\altaffiltext{5}{Universities Space Research Association.}

\begin{abstract}
We present a technique for optimal coaddition of image data for rapidly
varying sources, with specific application to gamma-ray burst (GRB)
afterglows.  Unweighted coaddition of rapidly fading afterglow lightcurve
data becomes counterproductive relatively quickly.  It is
better to stop coaddition of the data once noise dominates late exposures.
A better alternative is to optimally weight each exposure to maximize the
signal-to-noise ratio ($S/N$) of the final coadded image data.  By using
information about GRB lightcurves and image noise characteristics,
optimal image coaddition increases the probability of afterglow
detection and places the most stringent upper limits on non-detections.
For a temporal power law flux decay typical of GRB afterglows, optimal
coaddition has the greatest potential to improve the $S/N$ of afterglow
imaging data (relative to unweighted coaddition), when the decay rate
is high, the source count rate is low, and the background rate is high.
The optimal coaddition technique is demonstrated with applications to
{\em Swift} Ultraviolet/Optical Telescope (UVOT) data of several GRBs,
with and without detected afterglows.  \end{abstract}

\keywords{methods: data analysis; gamma-rays: bursts}

\section{Introduction}

The early detection of gamma-ray burst (GRB) optical afterglows
is crucial to fulfill several key science goals set forth by the
{\em Swift} Gamma-ray Burst Explorer mission \citep{gehrels04}.
The Ultraviolet/Optical telescope \citep[UVOT;][]{roming05a} onboard
\emph{Swift} was designed to capture these early ($\sim 1$ minute after
burst trigger) afterglows in order to both study the behavior of the
early optical afterglow and to provide accurate sub-arcsecond positions
for ground-based follow-up observations.  However, approximately $60\%$
of the GRBs detected by {\em Swift} lack an optical counterpart in rapidly
available UVOT data \citep{marshall07, romingmason06, mason05}, and more
than $40\%$ lack an optical or IR counterpart in data from any telescope.
This unexpectedly high fraction of optical non-detections highlights the
importance of ensuring that the available data are being used to their
full potential.

In particular, it is imperative that data are coadded in such a way
that the chance of revealing a faint detection is maximized or the most
stringent upper-limits are obtained.  The rapidly fading behavior of
GRBs ensures that coadding more data will eventually become detrimental
as the burst signal fades below the background level.  Upper limits have
often been reported via the Gamma-ray bursts Coordinates Network (GCN)
Circulars \citep{barthelmy95, barthelmy98} which, in our experience,
are often calculated using the coaddition of an arbitrary number of
unweighted images \citep[e.g.,][]{morgan05, roming05b}.  By continuing
to coadd capriciously, we run the risk of not reporting the most useful
information to the rest of the GRB community and may even overlook a faint
afterglow.  In order to best use the imaging data from \textit{Swift}, a
method for determining when to stop coadding burst data, or for optimally
coadding the data, must be implemented.

This paper describes the necessary criteria and the technique for
optimally coadding faint GRB afterglow imaging data, and for improving
detection limits when no afterglow has been detected in the individual
images.  Equations are derived that describe the signal-to-noise ratio
($S/N$) of a number of summed exposures for an object whose flux density
fades according to a simple power law decay
\begin{equation}
F_{\nu} \propto t^{\alpha},
\label{eq:powerlaw}
\end{equation}
where $\alpha < 0$ is the temporal decay index.  The $S/N$ of an optimally
coadded imaging data set depends on the decay index, the start and stop
times of each exposure, the measured counts in each exposure, and the
noise levels of each exposure.  For bursts which lack an optical afterglow
detection, however, the value of the crucial decay index parameter is
unknown, making it necessary to adopt assumptions in this case.

Observations with {\em Swift}'s X-ray Telescope \citep[XRT;][]{burrows05}
suggest a canonical GRB X-ray afterglow lightcurve, consisting of three
distinct power law segments with decay indices ranging from very steep
decay at early times ($-5 \lesssim \alpha \lesssim -3; t \lesssim 300$
s), followed by a shallow decay ($-0.8 \lesssim \alpha \lesssim -0.2$)
until $t \sim 10^4 - 10^5$\,s, and finally a decay of medium ($-1.5
\lesssim \alpha \lesssim -1$) steepness \citep{nousek06, zhang06}.
Optical and UV afterglows typically display an initial phase lasting
up to about 500\,s during which the lightcurve is slowly decaying
or even rising, and a longer steeper power law decay phase with a
temporal slope of about $\alpha = -0.9 \pm 0.4$ \citep{oates08}.
Flaring at early times, as is often seen in X-ray lightcurves
\citep[e.g.][]{nousek06}, is occasionally manifested in optical
lightcurves \citep[e.g.][]{holland02,jakobsson04,blustin06a,wei06,dai07}.
Optical and X-ray lightcurves may, but typically do not, track each other
closely.  We assume the simplest temporal decay model given by equation
(\ref{eq:powerlaw}) in our modeling of $S/N$, but it is straightforward
to generalize the technique to arbitrary lightcurves.

In \S\,\ref{sec:uvotData}, observations of GRB afterglows using the {\em
Swift} UVOT are described.  The optimal image data coaddition technique
is developed in \S\,\ref{sec:technique}.  The technique is tested with
simulated data in \S\,\ref{sec:simulations}, and applied to examples of
real UVOT afterglow data in \S\,\ref{sec:realData}.  A discussion and
summary are given in \S\,\ref{sec:summary}.

\section{{\em Swift} UVOT Observations of GRB Afterglows \label{sec:uvotData}}

The techniques described here apply generically to observations made by a
wide variety of telescopes and detectors, however, the UVOT telescope on
board the {\em Swift} satellite makes rapid and long-term observations
of almost all GRB fields identified by the {\em Swift} Burst Alert
Telescope \citep[BAT;][]{barthelmy05}.  For this reason the specific
examples used here will focus on data obtained with the UVOT.  In this
section we briefly describe the UVOT and the GRB afterglow data sets it
typically produces.

The {\em Swift} UVOT is a 30\,cm diameter telescope with a $17\times17$
arcminute field of view, with a detector sensitive to wavelengths
between approximately 1600 and 8000{\AA}, covered by seven photometric
filters labeled $uvw2, uvm2, uvm1, u, b, v$, and $white$ \citep{roming05a}.
The photometric calibration of the UVOT is described by \citet{poole08}.

The average time to the first UVOT observation of a GRB location after
the BAT trigger is 110s, barring constraints due to the positions of a
burst relative to the Sun, Earth, and Moon.  The automated GRB observing
sequence utilizes all seven filters, with the order and exposure times
determined with respect to the time of the burst.  The simulations
described in \S\,\ref{sec:simulations} are based on a typical set of
exposures in the UVOT $v$ band, up to about $10^{5}$\,s post-burst.
A normal set of $v$ band exposures would consist of a 400\,s ``finding
chart'' exposure shortly after slewing to the GRB position, a series of
short (10 to 20\,s) exposures, another finding chart exposure, another
series of short exposures, and finally a set of longer exposures (100
to 900\,s).  The sequence described here is typical for the majority of
{\em Swift} detected GRBs and is currently implemented for the detection
of new bursts.  Observations often continue to be made for several days
or weeks, depending on the flux of the afterglow seen by the XRT.  It is
very rare, however, for any significant flux to be detected at optical
or UV wavelengths at these late times; so late-time observations will
not be considered for the simulations in \S\,\ref{sec:simulations}.

GRB afterglows are detected in the first $white$ or $v$ observations
approximately $40\%$ of the time when those observations are made
less than 500\,s after the burst.  An analysis of the implications for the
nature of GRB afterglows based on the UVOT detection statistics is given
by \citet{roming06}.  Additional detections are sometimes made after the
coaddition of frames taken over a period of time.  In the next section,
we examine the effectiveness of standard unweighted frame coaddition
of GRB afterglow data, and describe a method for optimal coaddition of
imaging data of temporally varying sources.

\section{Optimal Coaddition of GRB Image Data \label{sec:technique}}

\subsection{Unweighted Coaddition}

Assuming a source with a lightcurve that follows a power law decay,
we derive an equation that gives an estimate of the $S/N$ in an
aperture surrounding the source for an unweighted sum of $n$ exposures.
The equation for the final $S/N$ depends on the initial source count
rate of the afterglow, the background count rate for each exposure, the
start and stop times of each exposure, and the temporal decay index of
the burst.  With this equation, one can calculate the exposure at which
the maximum $S/N$ occurs for a given burst if unweighted coaddition is
adopted.  We implicitly assume that the measurements are made in the same
aperture surrounding the same known source location in each of the images.
Summing the images and then measuring counts in an aperture in the summed
image is equivalent to summing the measurements from apertures in the
individual images; here we proceed as if individual aperture measurements
are being summed.

There are several sources of noise for a given detector.  Among them
are noise due to the signal itself ($N_{src}$), noise due to the
sky background ($N_{sky}$), dark noise ($N_{dark}$), and read noise
($N_{read}$).  Assuming all sources of noise are known to perfect accuracy
and are uncorrelated, the total noise is the sum in quadrature of these
four quantities:
\begin{equation}
N^2 = N^2_{src} + N^2_{sky} + N^2_{dark} + N^2_{read}.
\end{equation}
For the UVOT, the read noise is zero since it is a photon counting
device, and the dark noise is insignificant \citep{mason04}, and thus the
dominant contributors to the noise are the source and the sky background.
We assume that the source and sky counts are Poisson distributed, but
that the summed image counts are large enough that Gaussian statistics
apply, so that the total noise is simply the square root of the number
of counts:
\begin{equation}
  N^2 = \sum_i \left(C_{src,i} + C_{sky,i}\right),
  \label{eq:N2}
\end{equation}
where $C_{src,i}$ and $C_{sky,i}$ are the measured number of source and
sky background counts in the $i^{\rm th}$ exposure, respectively.
The signal-to-noise ratio of the summed image is thus given by
\begin{equation}
    S/N_{sum} = \frac{\sum_{i} C_{src,i}}
      {\sqrt{\sum_{i} \left(C_{src,i}+C_{sky,i}\right)}}.
      \label{eq:sn}
\end{equation}

The estimated total number of counts in an exposure from a given source
is the integral of the count rate $R(t)$ from the start to stop times of an
exposure.  If we approximate the sky background count rate ($R_{sky,i}$) as
constant during the $i^{th}$ exposure, the estimated sky counts in that
exposure are
\begin{eqnarray}
C_{sky,i} =  R_{sky,i} \left(t_{stop,i}-t_{start,i}\right),
\label{eq:bcsky}
\end{eqnarray}
where $t_{start,i}$ and $t_{stop,i}$ are the start and stop times of
the $i^{th}$ exposure after the initial burst trigger.

Using the assumption of a simple temporal power law decay for gamma-ray burst
afterglows, the model source count rate $R_{model}$ can be parameterized as
\begin{equation}
R_{model}(t) = R_{1} \left(\frac{t}{t_1}\right)^\alpha ,
\label{eq:rsrcfirst}
\end{equation}
where $\alpha$ is the temporal decay index and $R_{1}$ is the initial
source count rate, defined to be the average count rate of the $1^{st}$
exposure given by
\begin{equation}
R_{1} = \frac{C_{src,1}}{\left(t_{stop,1}-t_{start,1}\right)}\,.
\label{eq:r0}
\end{equation}
The parameter $t_1$ is thus the weighted midpoint of the exposure at
which time the count rate is $R_{1}$.  Integrating (\ref{eq:rsrcfirst}),
the model number of source counts in the $i^{th}$ exposure is given by
\begin{eqnarray}
C_{model,i}  = 
	\left[
	\begin{array}{ll}
	  \frac{R_1}{t_1^\alpha \left(\alpha+1 \right)}
            \left(t_{stop,i}^{\alpha+1} - t_{start,i}^{\alpha+1} \right)
            & \quad \alpha \neq -1 \\
	  \frac{R_1}{t_1^\alpha} \ln 
            \left(\frac{t_{stop,i}}{t_{start,i}}\right)
            & \quad \alpha = -1 \\
         \end{array} \right]
\label{eq:bcsrc}
\end{eqnarray}
For the remainder of the derivation, we shall assume for simplicity that
$\alpha$ is never exactly -1.  Inserting (\ref{eq:r0}) into (\ref{eq:bcsrc})
for $i=1$ and solving for $t_1$ gives
\begin{equation}
t_1 = \left[\frac{ \left(t_{stop,1}^{\alpha+1} - t_{start,1}^{\alpha+1} \right)}{\left(\alpha+1 \right) \left(t_{stop,1}-t_{start,1}\right)} \right]^\frac{1}{\alpha}\,.
\label{eq:t0}
\end{equation}
Inserting (\ref{eq:t0}) into (\ref{eq:bcsrc}) gives
\begin{equation}
  C_{model,i} = R_1  \left(t_{stop,1}-t_{start,1}\right) 
    \frac{\left(t_{stop,i}^{\alpha+1} - t_{start,i}^{\alpha+1} \right)}
    {\left(t_{stop,1}^{\alpha+1} - t_{start,1}^{\alpha+1} \right)}\,.
  \label{eq:csrc}
\end{equation}

The time since the burst associated with the coadded data is not well
defined, since it involves many different time intervals, each sampling a
different portion of a light curve.  However, a reasonable definition for
a characteristic post burst time, $\bar{t}$, is the count weighted average 
time of the observations. In other words, $\bar{t}$ is the average arrival
time of the counts.  Because the count rate is expected to change across
an exposure, the observing times are weighted by the integrated model
count rate across each exposure window, rather than weighting them with
the measured counts.  Thus we arrive at the characteristic time
\begin{eqnarray}
  \bar{t} & = & \frac{\sum_{i}^{n}\int_{t_{start,i}}^{t_{stop,i}}tR_{model}dt}
      {\sum_{i}^{n}\int_{t_{start,i}}^{t_{stop,i}}R_{model}dt}
      \label{eq:t_barA}\\
\bigskip
  & = & \left(\frac{\alpha + 1}{\alpha + 2}\right)
      \frac{\sum_{i}^{n}\left(t_{stop,i}^{\alpha + 2}
          - t_{start,i}^{\alpha + 2}\right)}
      {\sum_{i}^{n}\,\left(t_{stop,i}^{\alpha + 1}
          - t_{start,i}^{\alpha + 1}\right)} \,,
      \label{eq:t_barB}
\end{eqnarray}
which is independent of the initial count rate.

If the temporal decay model is a good approximation to the real lightcurve,
the $S/N$ for a summed set of exposures can be predicted as
\begin{eqnarray}
S/N_{pred} = \frac{R_1  \left(t_{stop,1}-t_{start,1}\right)
   \displaystyle\sum_{i=1}^n
   \frac{\left(t_{stop,i}^{\alpha+1} - t_{start,i}^{\alpha+1} \right)}
   {\left(t_{stop,1}^{\alpha+1} - t_{start,1}^{\alpha+1} \right)} }
   {\sqrt{R_1  \left(t_{stop,1}-t_{start,1}\right) 
   \displaystyle\sum_{i=1}^n
   \frac{\left(t_{stop,i}^{\alpha+1} - t_{start,i}^{\alpha+1} \right)}
   {\left(t_{stop,1}^{\alpha+1} - t_{start,1}^{\alpha+1} \right)}
   + \displaystyle\sum_{i=1}^n R_{sky,i}
   \left(t_{stop,i}-t_{start,i}\right)}}\,.
\label{eq:snfinal}
\end{eqnarray}
With this equation, one can predict the signal-to-noise for a summed
image of $n$ exposures by simply knowing the start and stop times of each
exposure (relative to the time of burst) and measuring both the initial
source count rate and the background count rates for each exposure.
In typical sets of observations, the $S/N$ will initially increase with
the coaddition of more frames, but as the count rate drops the noise
will increase faster than the signal, until eventually the $S/N$ will
reach a peak value, then begin to fall.  By modeling the S/N, one can
determine the amount of data to coadd to achieve the most significant
detection for unweighted data.

\subsection{Weighted Exposures}

While equation (\ref{eq:snfinal}) can be used to predict a peak $S/N$ in
unweighted coadded data, the significance of a source detection in
a series of images can be raised by optimally weighting the data in
each exposure.  Additionally, with optimal weighting, the $S/N$ will not
drop as the source counts become less significant.  The derivation
of the optimal weighting scheme parallels in part the arguments given by
\citet{horne86} for the optimal extraction of spectroscopic data.
In that case, the spatial profile of a spectrum at a given wavelength
was used to provide independent estimates, at each pixel, of the total
source counts in the spectrum.  Each estimate was weighted in such
a way that the average value gives the minimum variance in the
the total source counts.  In the case of GRBs, the temporal lightcurve
takes the place of the spatial spectrum profile, and each
observation is the synonym of a pixel in the spectrum profile.

Let $P_{i}$ be the probability that a detected photon from a GRB
afterglow is recorded in frame $i$.  The sum of the probabilities
over all the $n$ frames is normalized to unity
\begin{eqnarray}
  \displaystyle\sum_{i=1}^n P_{i} = 1\,.
  \label{eq:probSum}
\end{eqnarray}
The probability for frame $i$ is given by the expected GRB afterglow
counts in that frame, $C_{model,i}$ (given, for example by equation
(\ref{eq:csrc})), divided by the total number of expected counts, which
is the sum over the expected counts in all of the frames
\begin{eqnarray}
  P_{i} = \frac{C_{model,i}}{\sum_{i}^n C_{model,i}}\,.
  \label{eq:P_i}
\end{eqnarray}
The probability function is the normalized model GRB afterglow lightcurve.
We assume here that the lightcurve is known, but we will discuss
the case in which the lightcurve is uncertain in \S\,\ref{sec:realData}.

Given the normalized lightcurve, an estimate of the total number of
source counts can be made for each frame, by dividing the source
counts by the probability function for that frame
\begin{eqnarray}
  C_{tot,i} \approx C_{src,i}/P_{i}\,.
  \label{eq:C_toti}
\end{eqnarray}
The average value of all of the estimates is then a linear and unbiased
estimator of the total number of source counts \citep{horne86}.
Including a weighting factor $w_{i}$, for each frame, the estimator of
the total source counts $C_{tot}$ is
\begin{eqnarray}
  C_{tot} = \frac{\sum_{i}^{n} w_{i}C_{src,i}/P_{i}}
            {\sum_{i}^n w_{i}}\,.
  \label{eq:C_totSum1}
\end{eqnarray}
The total count estimate is optimal when the variance on the estimator
is minimized.  The variance of equation (\ref{eq:C_totSum1}), assuming
equation (\ref{eq:N2}),  is
\begin{eqnarray}
  V(C_{tot}) = \frac{1}{\left(\sum_{i}^n w_{i}\right)^2}
      \displaystyle\sum_{i}^{n} w_{i}^2
      \left(C_{src,i} + C_{sky,i}\right)/P_{i}^2\,,
  \label{eq:var_C_tot}
\end{eqnarray}
which is minimized when
\begin{eqnarray}
  w_{i} = \frac{P_{i}^{2}}{C_{src,i} + C_{sky,i}}\,,
  \label{eq:weight}
\end{eqnarray}
modulo a multiplicative constant, so that equation (\ref{eq:C_totSum1})
becomes
\begin{eqnarray}
  C_{tot} = \frac{\sum_{i}^{n} P_{i} C_{src,i}
      / \left(C_{src,i} + C_{sky,i}\right)}
      {\sum_{i}^{n} P_{i}^{2}
      / \left(C_{src,i} + C_{sky,i}\right) }\,.
  \label{eq:C_totSum2}
\end{eqnarray}
For a constant source, equation (\ref{eq:C_totSum2}) reduces to the
familiar inverse variance weighting.  The optimal weighting factor
(\ref{eq:weight}) not only accounts for differences in the variances of
each frame, but weights each frame differently depending on the number of
source counts expected in the frames.  The $S/N$ of the coadded weighted
image data is
\begin{eqnarray}
  S/N_{sum,weight} = \frac{\sum_{i}^{n} P_{i} C_{src,i}
      / \left(C_{src,i} + C_{sky,i}\right)}
      {\left[\sum_{i}^{n} P_{i}^{2}
      / \left(C_{src,i} + C_{sky,i}\right)\right]^{1/2}}\,.
  \label{eq:sn_wt1}
\end{eqnarray}
In contrast to the unweighted $S/N$ given in equation (\ref{eq:sn}),
the weighted sum $S/N$ runs little risk of decreasing as the count rate
drops, if the lightcurve model is sufficiently accurate.  Frames with few
expected counts contribute little to either the final signal or noise.
The coaddition process does not need to stop at a particular frame to
achieve a maximum $S/N$ -- statistically, the maximum $S/N$ will always
be reached with the coaddition of all of the frames.

The expected number of counts in each frame is found from the lightcurve
model.  Assuming again that GRB afterglow lightcurves
can be described by a single power law, as in equation (\ref{eq:csrc}),
and using equation (\ref{eq:P_i}) for the probabilities $P_{i}$, the
optimally weighted total number of afterglow counts in a set of exposures is
\begin{eqnarray}
  C_{tot} = \left[
      \sum_{i}^n\left(t_{stop,i}^{\alpha+1}-t_{start,i}^{\alpha+1}\right)
      \right]
      \frac{\sum_{i}^{n}
      \left(t_{stop,i}^{\alpha+1}-t_{start,i}^{\alpha+1}\right) C_{src,i}
      / \left(C_{src,i} + C_{sky,i}\right)}
      {\sum_{i}^{n}
      \left(t_{stop,i}^{\alpha+1}-t_{start,i}^{\alpha+1}\right)^{2}
      / \left(C_{src,i} + C_{sky,i}\right) }\,,
  \label{eq:C_totSum4}
\end{eqnarray}
and the $S/N$ is
\begin{eqnarray}
  S/N_{sum,weight} = \frac{\sum_{i}^{n}
      \left(t_{stop,i}^{\alpha+1}-t_{start,i}^{\alpha+1}\right) C_{src,i}
      / \left(C_{src,i} + C_{sky,i}\right)}
      {\left[\sum_{i}^{n}
      \left(t_{stop,i}^{\alpha+1}-t_{start,i}^{\alpha+1}\right)^{2}
      / \left(C_{src,i} + C_{sky,i}\right) \right]^{1/2}}\,,
  \label{eq:sn_wt}
\end{eqnarray}
This shows that $C_{tot}$ can be determined independently
of the initial model count rate $R_{1}$, since the equation instead
contains the decay slope and observing times.  The total count estimate
does not correspond to a particular count rate, because the count rate
changes over time.  However, the initial estimated count rate can be found by
setting the weighted total counts equal to the sum of the model counts
in equation (\ref{eq:csrc}), and solving for $R_{1}$.  Then the count
rate at any moment can be estimated from the lightcurve model in equation
(\ref{eq:rsrcfirst}).

The characteristic time since the burst of the observations can be defined
in a manner similar to equation (\ref{eq:t_barA}), except that in this case each
exposure is further weighted by its optimal weight, $w_i$,
\begin{eqnarray}
  \bar{t} = \frac{\sum_{i}^{n}w_{i}\int_{t_{start,i}}^{t_{stop,i}}tR_{model}dt}
      {\sum_{i}^{n}w_{i}\int_{t_{start,i}}^{t_{stop,i}}R_{model}dt}\,.
  \label{eq:t_barWtA}
\end{eqnarray}
For the optimal weights defined by equation (\ref{eq:weight}), the
characteristic time is
\begin{eqnarray}
  \bar{t} = \left(\frac{\alpha + 1}{\alpha + 2}\right)
      \frac{\sum_{i}^{n}\left(t_{stop,i}^{\alpha + 2} 
          - t_{start,i}^{\alpha + 2}\right)
          \left(t_{stop,i}^{\alpha + 1} - t_{start,i}^{\alpha + 1}\right)^{2}
          / \left(C_{src,i} + C_{sky,i}\right)}
      {\sum_{i}^{n}\,\left(t_{stop,i}^{\alpha + 1} 
          - t_{start,i}^{\alpha + 1}\right)^{3} 
          / \left(C_{src,i} + C_{sky,i}\right)}\,,
  \label{eq:t_barWtB}
\end{eqnarray}
which is independent of the initial count rate.

\subsection{Detection Limits \label{sec:detection}}

When an afterglow is not detected in any single image, coaddition of the
image data may improve the final $S/N$ enough to provide a detection.
Even when the final coadded image does not yield a detection, optimal
coaddition can be used to place stronger detection limits on the dataset,
than would result from unweighted coaddition.  If the lightcurve
shape can be assumed, the source counts $C_{src,i}$, can be replaced
by the lightcurve model $C_{model,i}$ in equations (\ref{eq:sn}) and
(\ref{eq:sn_wt1}), and the equations set to a minimum $S/N$ (e.g. 3)
required for a detection, then solved for the parameters of the lightcurve
model.  For the case of a power law decay with a given index
(equation (\ref{eq:csrc})), this will yield the maximum initial average
count rate, $R_{1,max}$ that could have occurred without producing a
detection in the coadded data.  In other words, given a non-detection in
the coadded data, $R_{1,max}$ is the limiting measured count rate of the
source during the time of the first exposure.

For unweighted data, the expression for $R_{1,max}$ can be derived by
solving equation (\ref{eq:snfinal}) for $R_{1,max}$.  The solution
is quite simple in that case, since the equation is a quadratic.
The unweighted maximum initial count rate may be either lower or higher
(better or worse respectively) than the detection limit of the first image
by itself, depending on how many unweighted noise-dominated exposures
are coadded.

When the data are optimally coadded, equation (\ref{eq:sn_wt1}) applies,
with $C_{src,i}$ replaced with $C_{model,i}$.  Assuming a power law
decay model for the lightcurve, the $S/N$ is described by equation
(\ref{eq:sn_wt}); using $C_{model,i}$ in place of $C_{src,i}$, that
equation cannot be solved analytically for $R_{1,max}$, but it is not
difficult to solve numerically.  Initial (but slightly over-estimated)
guesses for $R_{1,max}$ can be made by assuming that the source counts
in the denominator of equation (\ref{eq:sn_wt}) are negligible compared
to the background counts, which yields an equation that can be solved
analytically for $R_{1,max}$,
\begin{eqnarray}
R_{1,max} \approx \frac{S/N_{min}}{\left(t_{stop,1}-t_{start,1}\right)}
   \left[ \displaystyle\sum_{i=1}^n
   \frac{\left(t_{stop,i}^{\alpha+1} - t_{start,i}^{\alpha+1} \right)^{2}}
   {\left(t_{stop,1}^{\alpha+1} - t_{start,1}^{\alpha+1} \right)^{2}
   C_{sky,i}} \right]^{-1/2} \,,
\end{eqnarray}
where $S/N_{min}$ is the minimum $S/N$ required for the detection of
a source.  By utilizing the optimally coadded data, the initial average
count rate limit will always be more sensitive than the limit of the
first image by itself.  Statistically, the optimally weighted $R_{1,max}$
will be more sensitive than the value calculated from the unweighted data.

\section{Tests With Simulated GRB Afterglow Data \label{sec:simulations}}

The final coadded $S/N$ of GRB afterglow data depends upon the observing
times, the initial afterglow count rate, the temporal decay slope,
the background count rate, and the count rates in each of the images.
To test how the quality of coadded data depends on these parameters, we
have constructed simulated GRB afterglow lightcurve data, and applied
both unweighted and optimally weighted coaddition.  While focusing on
typical UVOT parameters, the qualitative results of the simulations
apply generally to most GRB afterglow observations.

For each simulation, the decay slope $\alpha$, the initial afterglow count
rate $R_{1}$, and the background count rate $C_{sky}$ (assumed constant
for simplicity) were specified.  The observing times were kept the same
in each case, and are representative of a typical UVOT observing sequence
in the $v$ band (See \S\,\ref{sec:uvotData}).  In each observing time
interval, the lightcurve model was integrated over the exposure times
to find the expected number of counts, which was used as a Poisson mean.
The background count rate was also integrated over the exposure times
to find the Poisson mean background counts for each time interval.
The simulated detected counts from both the afterglow and background were
drawn from Poisson distributions, given their respective mean values.
The sum of the detected afterglow and background counts is the total
simulated observed counts.  The mean background counts (assumed to
be accurately measurable) were subtracted from the total counts to
give the simulated measured afterglow counts.  Using these values, the
coadded $S/N$ at each time interval was calculated according to equation
(\ref{eq:sn}) for the unweighted case, and equation (\ref{eq:sn_wt})
for the optimally weighted case.

\begin{figure}
\plottwo{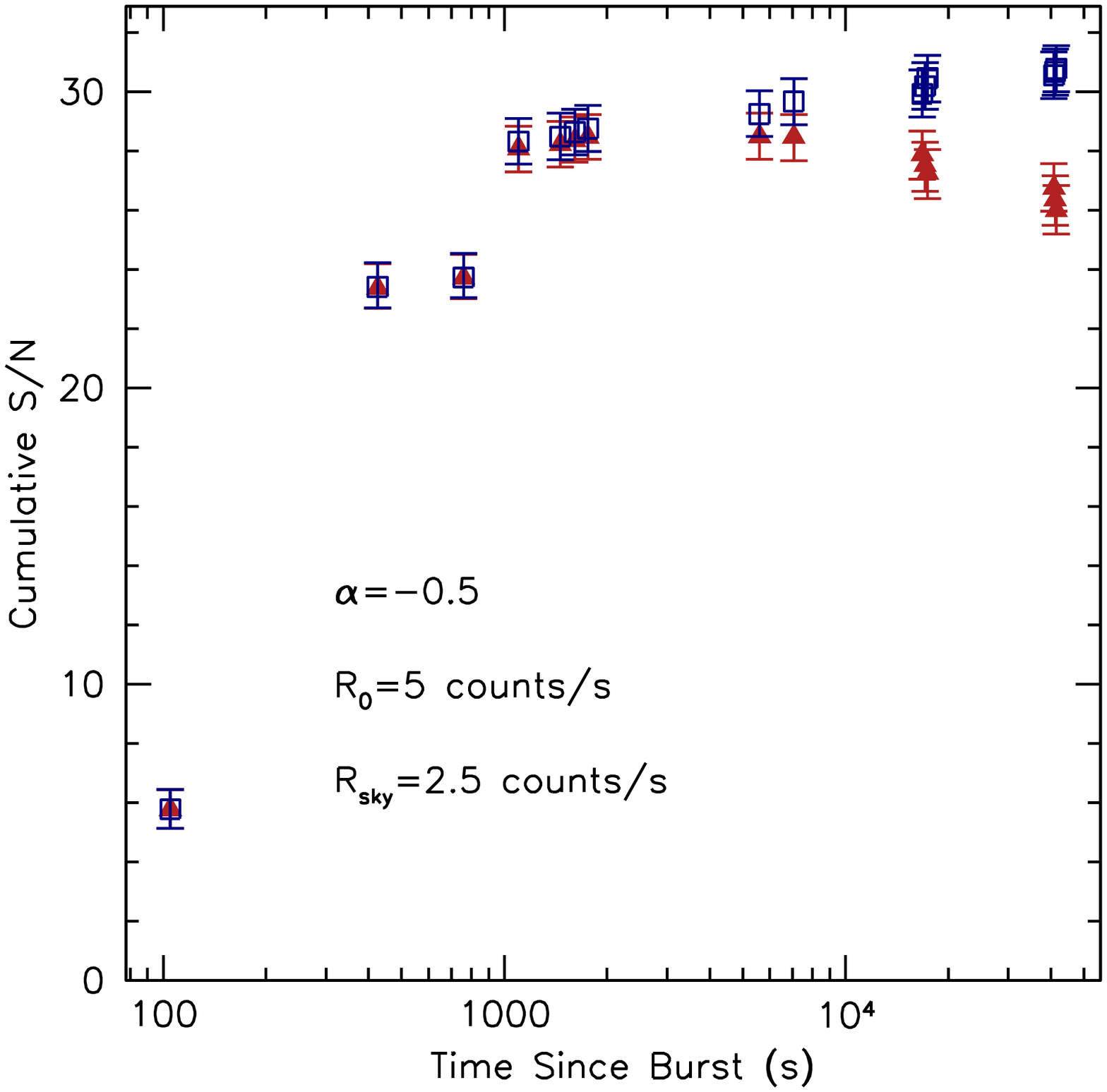}{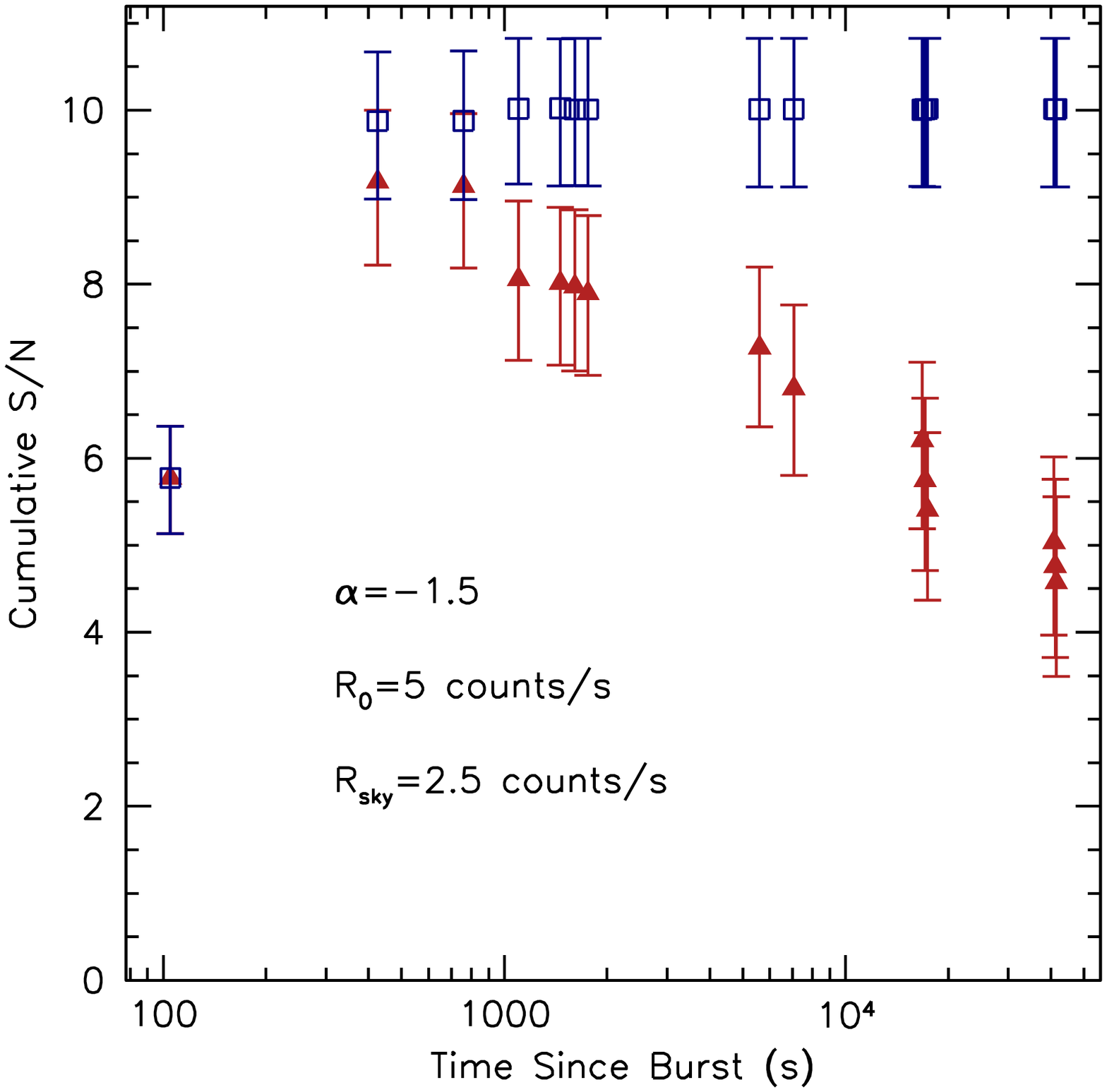}
\caption{Cumulative $S/N$ as a function of time since the burst, for
  coadded simulated GRB afterglow data, with a shallow decay slope
  (left) and a steep decay slope (right).  The temporal decay slopes
  $\alpha$, and the initial count rate $R_{1}$, of the lightcurve models
  are given, along with the background count rate $R_{sky}$. Each point
  corresponds to an observation in a relatively short time interval.
  Unweighted coadded data is shown by red triangles; optimally weighted
  coadded data is shown by blue squares.  The error bars span the $68.3\%$
  confidence limits in the results of 1000 simulations.}
  \label{fig:simDecaySlope}
\end{figure}

Optical temporal decay slopes for GRB afterglows are usually observed to
be in the range $\alpha = -0.5$ to $-2.0$, with an average of about $-0.9$
\citep{oates08}.  To see how the effectiveness of optimal coaddition
is affected by different decay slopes, data were coadded for simulated
lightcurves with a range of decay slopes, while the other parameters
were held constant.  Figure\,\ref{fig:simDecaySlope} shows the cumulative
$S/N$ after coadding the data through each observing interval, using both
unweighted, and optimally weighted coaddition, for two different values
of the decay slope.  Error bars show the $68.3\%$ (nominal $1\sigma$)
confidence limits from the results of 1000 simulations.  In each case,
the unweighted $S/N$ reaches a peak value, then decreases as more data
are coadded.  This happens because the noise dominates the observations to
a greater extent with each successive observation.  In contrast, the $S/N$
of the optimally weighted data rises, then flattens to a nearly constant
value at long post-burst times.  The late-time data contributes very
little signal or noise, so the $S/N$ values remain almost unchanged.
The effect is the same, regardless of the value of the decay slope,
but the relative difference between the final weighted and unweighted
$S/N$ is greatest for the data with the steepest decay slope.  When the
lightcurves are steep, the afterglow fades below the noise level more
quickly, so the unweighted $S/N$ drops more rapidly, while the optimally
weighted $S/N$ flattens out sooner.

\begin{figure}
\plottwo{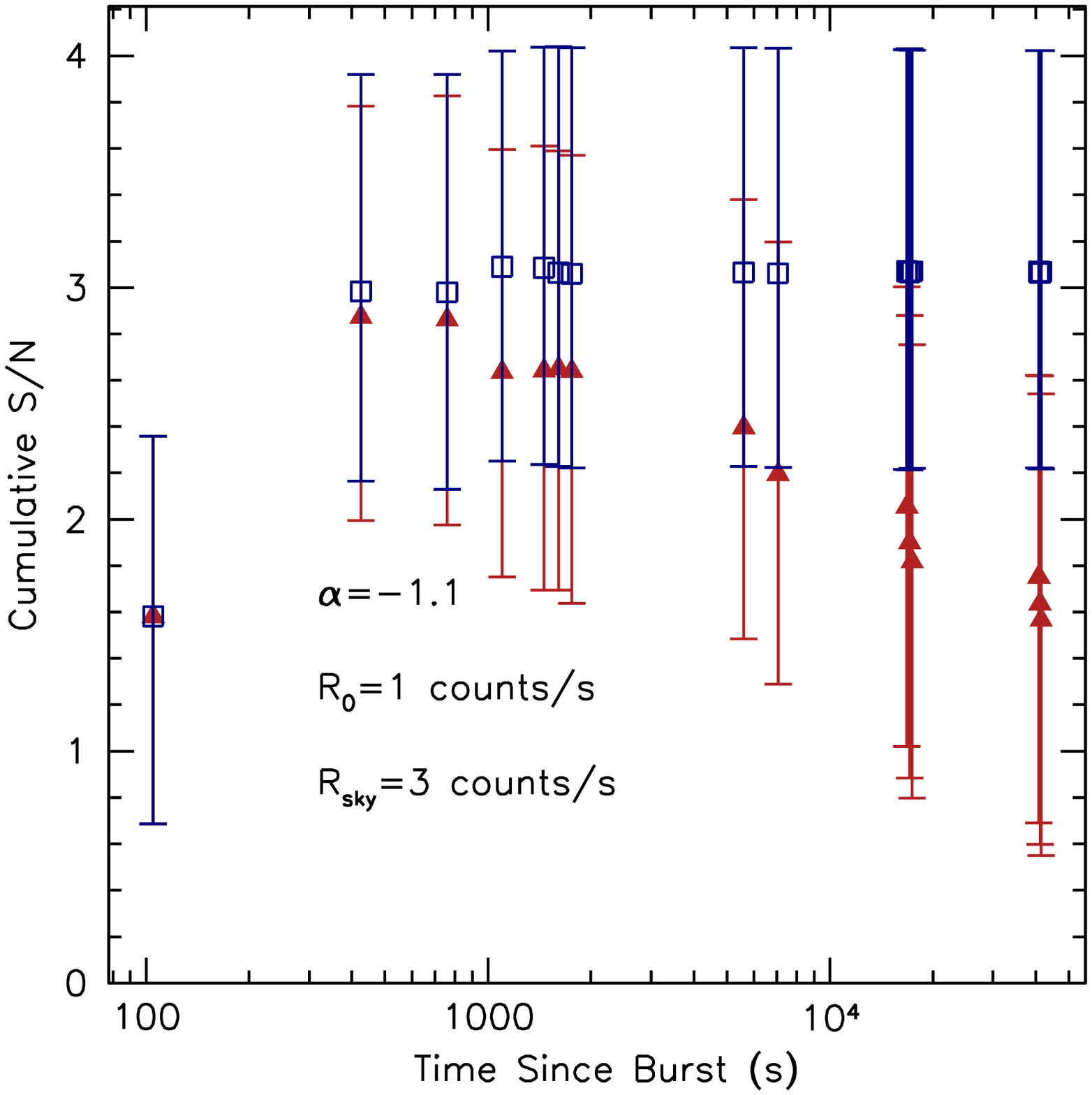}{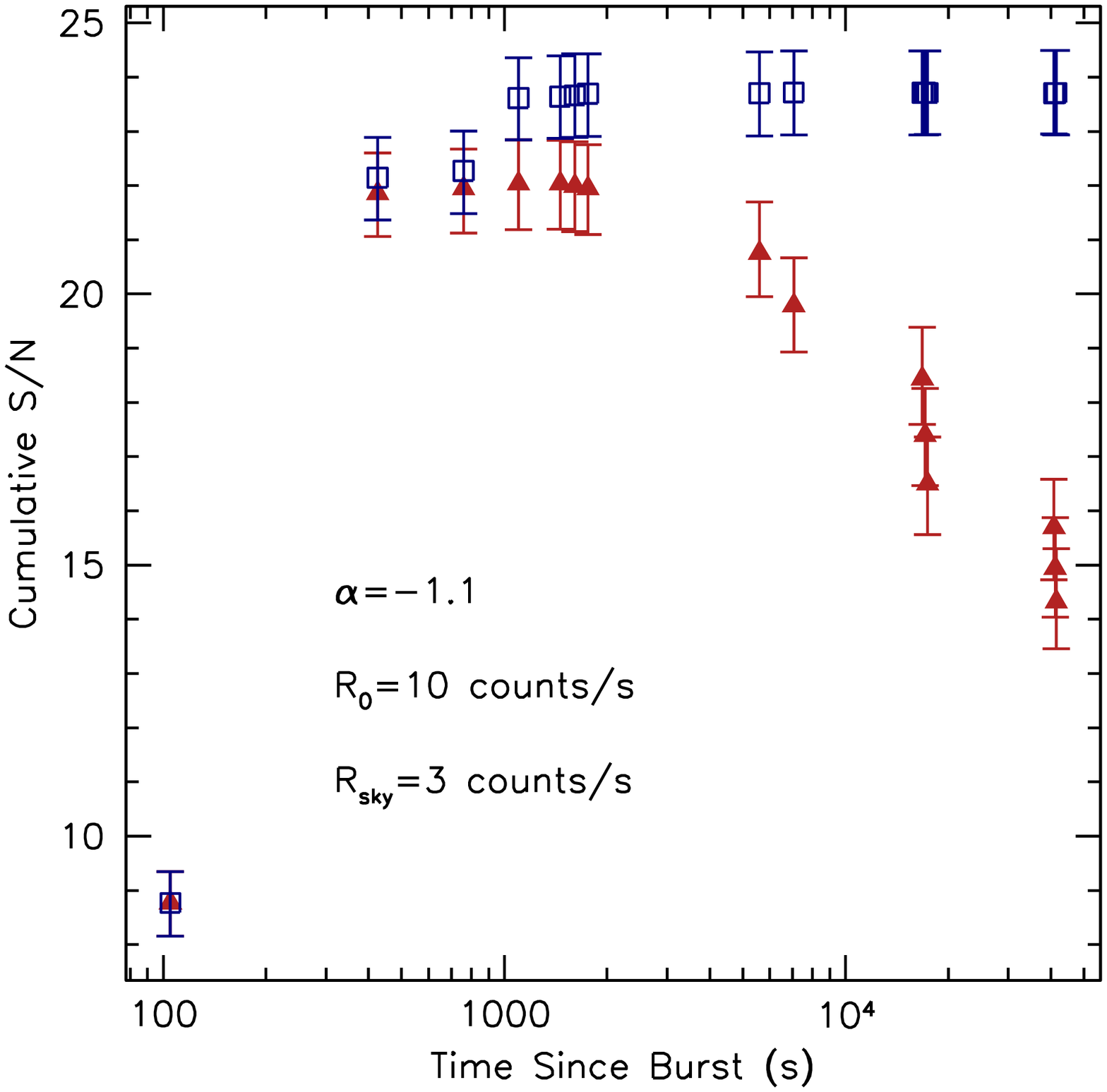}
\caption{Cumulative $S/N$ as a function of time since the burst, for
  coadded simulated GRB afterglow data, with a relatively low initial
  count rate (left) and high initial count rate (right).  The temporal
  decay slope $\alpha$, and the initial count rates $R_{1}$, of the
  lightcurve models are given, along with the background count rate
  $R_{sky}$. Each point corresponds to an observation in a relatively
  short time interval.  Unweighted coadded data is shown by red triangles;
  optimally weighted coadded data is shown by blue squares.  The error
  bars span the $68.3\%$ confidence limits in the results of 1000
  simulations.}
  \label{fig:simSourceRate}
\end{figure}

Figure\,\ref{fig:simSourceRate} shows how the coadded $S/N$ is affected
by differences in the initial count rate; the decay slope and background
count rates were fixed at typical values.  The results show that the
final weighted $S/N$ is significantly better than the unweighted $S/N$
in all cases, and that optimal coaddition has the greatest potential to
improve the final $S/N$ when the count rate is low.  These statements
are true in a statistical sense, but at low count rate, there is no
guarantee that the cumulative weighted $S/N$ at any observing time will
exceed the unweighted value, as the dispersion of the simulations shows.
Nonetheless, as shown by Fig.\,\ref{fig:simSourceRate}, in typical cases
optimal coaddition is more likely to result in a significant ($S/N > 3$)
detection, than when using unweighted coaddition.

\begin{figure}
\plottwo{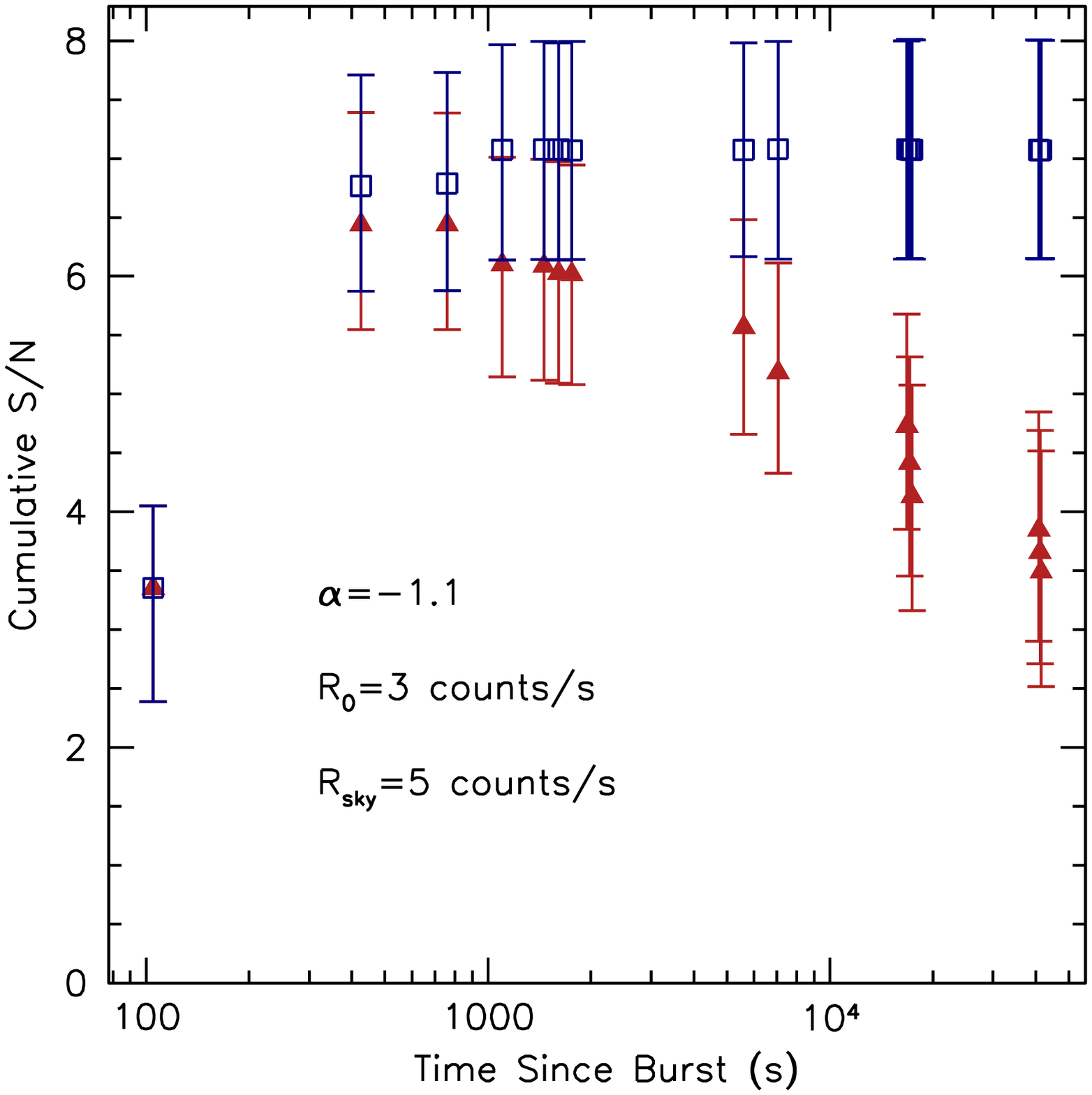}{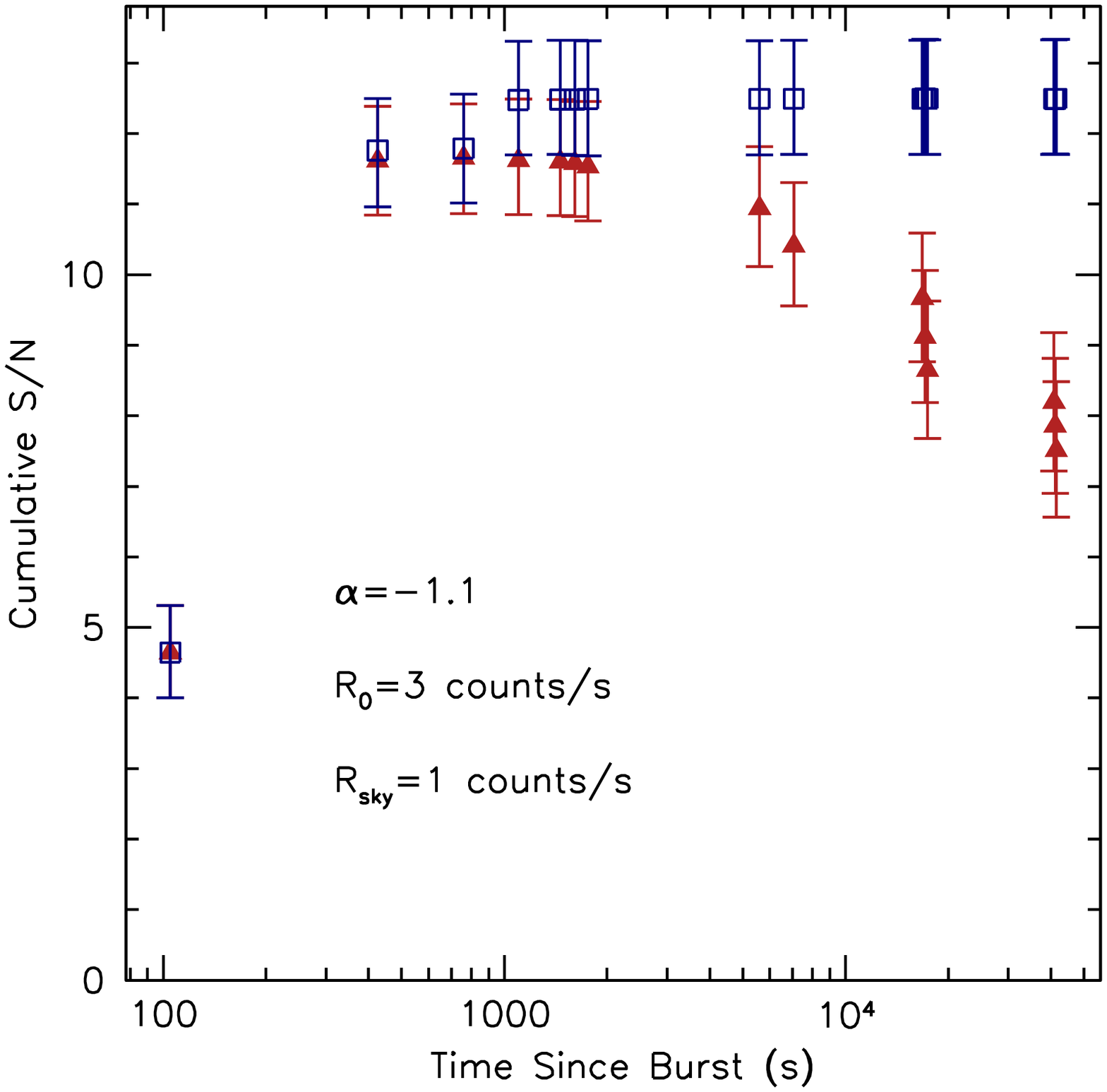}
\caption{Cumulative $S/N$ as a function of time since the burst, for
  coadded simulated GRB afterglow data, with a high background count rate
  (left) and a low background count rate (right).  The temporal decay
  slope $\alpha$, and the initial count rate $R_{1}$, of the lightcurve
  models are given, along with the background count rates $R_{sky}$.  Each
  point corresponds to an observation in a relatively short time interval.
  Unweighted coadded data is shown by red triangles; optimally weighted
  coadded data is shown by blue squares.  The error bars span the $68.3\%$
  confidence limits in the results of 1000 simulations.}
  \label{fig:simBkgRate}
\end{figure}

The background count rate clearly affects the detectability of fading
afterglows.  The background in UVOT images (although low compared
to ground-based telescope images) can change greatly from filter to
filter, and even among images using the same filter, depending on the
directions of the Sun, Earth, and Moon.  Figure\,\ref{fig:simBkgRate}
shows that optimal coaddition improves the $S/N$ in the presence of
a wide range of background rates (with decay slope and initial count
rate fixed in the simulations), particularly when the background rate
is high.  When background count rates are very low, as is often the case
with X-ray telescopes, there is little benefit from optimal coaddition.
For most optical and infra-red detectors, the background is usually the
dominant noise factor, so optimal coaddition is worth the minimal
implementation effort.

\begin{figure}
\plotone{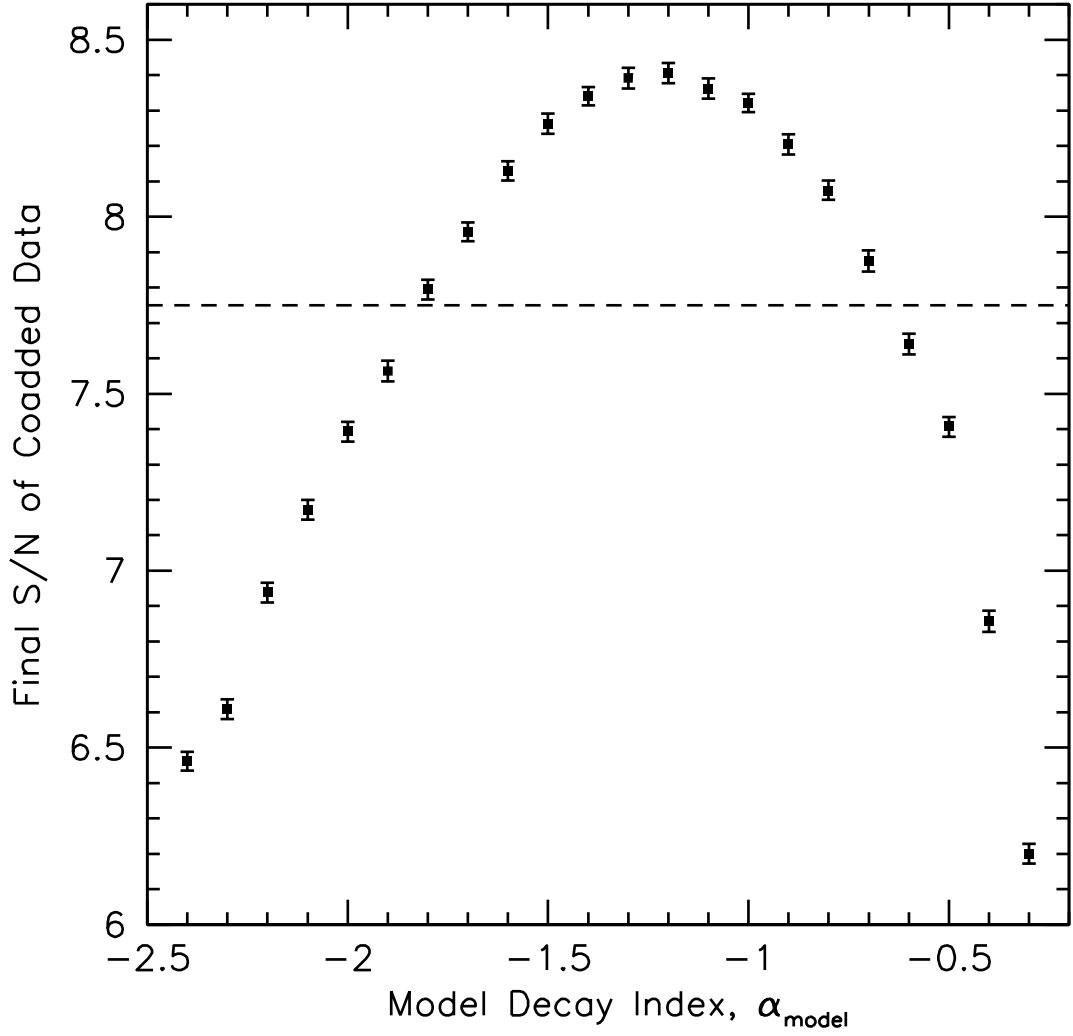}
\caption{Final $S/N$ in the optimally coadded data as a function of
  the model temporal decay index $\alpha_{model}$, for simulated
  GRB afterglow data.  The true decay index is $-1.2$, and the simulations
  used an initial count rate of 3\,${\rm cts\, s^{-1}}$, and a background
  count rate of 2.5\,${\rm cts\, s^{-1}}$.  The points give the median
  values from 1000 simulations, and the error bars show the uncertainty
  on the median value (not the dispersion in the simulations).  The
  dashed line shows the maximum $S/N$ from unweighted coadded data 
  averaged over the simulations.}
  \label{fig:finalSN}
\end{figure}

Often the decay slope is unknown, or is not known with great accuracy,
so it is of interest to know how the final $S/N$ depends upon the decay
slope used in the optimal coaddition.  Simulations were made using typical
values of the true decay slope $\alpha = -1.2$, the initial afterglow
count rate $R_{1} = 3\,{\rm counts\, s^{-1}}$, and the background count
rate $C_{sky} = 2.5\,{\rm counts\,s^{-1}}$.  The model decay slope
$\alpha_{model}$ (not necessarily the same as the true slope), used for
the optimal coaddition was varied over a wide range of values.  The median
final $S/N$ in the coadded data simulations were calculated for each value
of $\alpha_{model}$.  The results are shown in Fig.\,\ref{fig:finalSN}.
As expected, the final $S/N$ is maximized when the model decay slope
closely matches the true slope.  Figure\,\ref{fig:finalSN} shows that,
at least for typical cases, there is some leeway in the value of the
model decay slope; differences of up to a few tenths in the slope change
the final $S/N$ only a small amount.  However, the model slope value
cannot be arbitrarily different from the true value, or the final $S/N$
can actually drop well below the maximum $S/N$ in the unweighted case
(shown by a dashed line in Fig.\,\ref{fig:finalSN}).

The results shown in Fig.\,\ref{fig:finalSN} suggest a way to estimate
the decay slope when it is not known.  Optimal data coaddition can be
repeated for a data set, for a range of values of the model decay slope.
The value that maximizes the final $S/N$ is most likely to come closest
to the true decay slope.  For the example given here, we calculated the
value of the decay slope that maximized the $S/N$ in the optimally coadded
data, for each simulated light curve.  The mean value of the decay slope
was $\alpha = -1.20$ with a standard deviation among the simulations
of $\sigma(\alpha)=0.19$.  The decay slope estimation technique works
well even in cases for which an afterglow is detected only after optimal
coaddition.  For example, in simulations where the afterglow decay slope
is $\alpha=-0.9$ (the average given by \citet{oates08}), the initial count
rate is $R_{1} = 0.9\,{\rm counts\, s^{-1}}$, and the background count
rate $C_{sky} = 5.0\,{\rm counts\,s^{-1}}$, the optimally coadded $S/N$
is just over 3, while the estimated value of the decay slope is $\alpha =
-0.89 \pm 0.44$.  Thus, this technique can be used to calculate reasonable
estimates of the decay slope, even in low $S/N$ circumstances.

As mentioned earlier, X-ray and optical lightcurves usually do not
track each other very well, particularly at early times.  There are
exceptions to this \citep[e.g.][]{blustin06a,grupe07}, but usually the
X-ray decay slope is steeper than the optical slope \citep{oates08},
so at best the X-ray slope can be used as a lower limit for the optical
slope estimate.  When the optical slope is unknown, the best approach
for optimal coaddition is probably to assume a range of reasonable decay
slopes, bracketed at the steep end by the X-ray slope, and to calculate
limiting magnitudes in each case.

\clearpage
\section{Application to Swift GRB Image Data \label{sec:realData}}

\subsection{Optically Bright Bursts}

To demonstrate the effect of optimal coaddition on real data, imaging
data from the bright well-sampled optical afterglow of GRB\,050525A were
coadded using the weighted and unweighted methods.  The GRB was
discovered by the {\em Swift} BAT, and the spacecraft slewed to the
source immediately \citep{band05}.  The UVOT began observations
about 65s after the burst, and a bright afterglow was detected in images
taken with all seven UVOT filters \citep{holland05}.  UVOT imaging data
of the afterglow field were collected up to about $1.2\times10^{6}$s
after the burst; we have used data only up to a few $\times 10^{5}$s
after the burst, since the much later time data are entirely consistent with
no source flux and add nothing to this example.  Full analysis
of the {\em Swift} data was presented by \citet{blustin06a}.  Here
we show how the $S/N$ of UVOT data from a well-detected afterglow 
can depend strongly on how the data are coadded.

\begin{figure}
\plotone{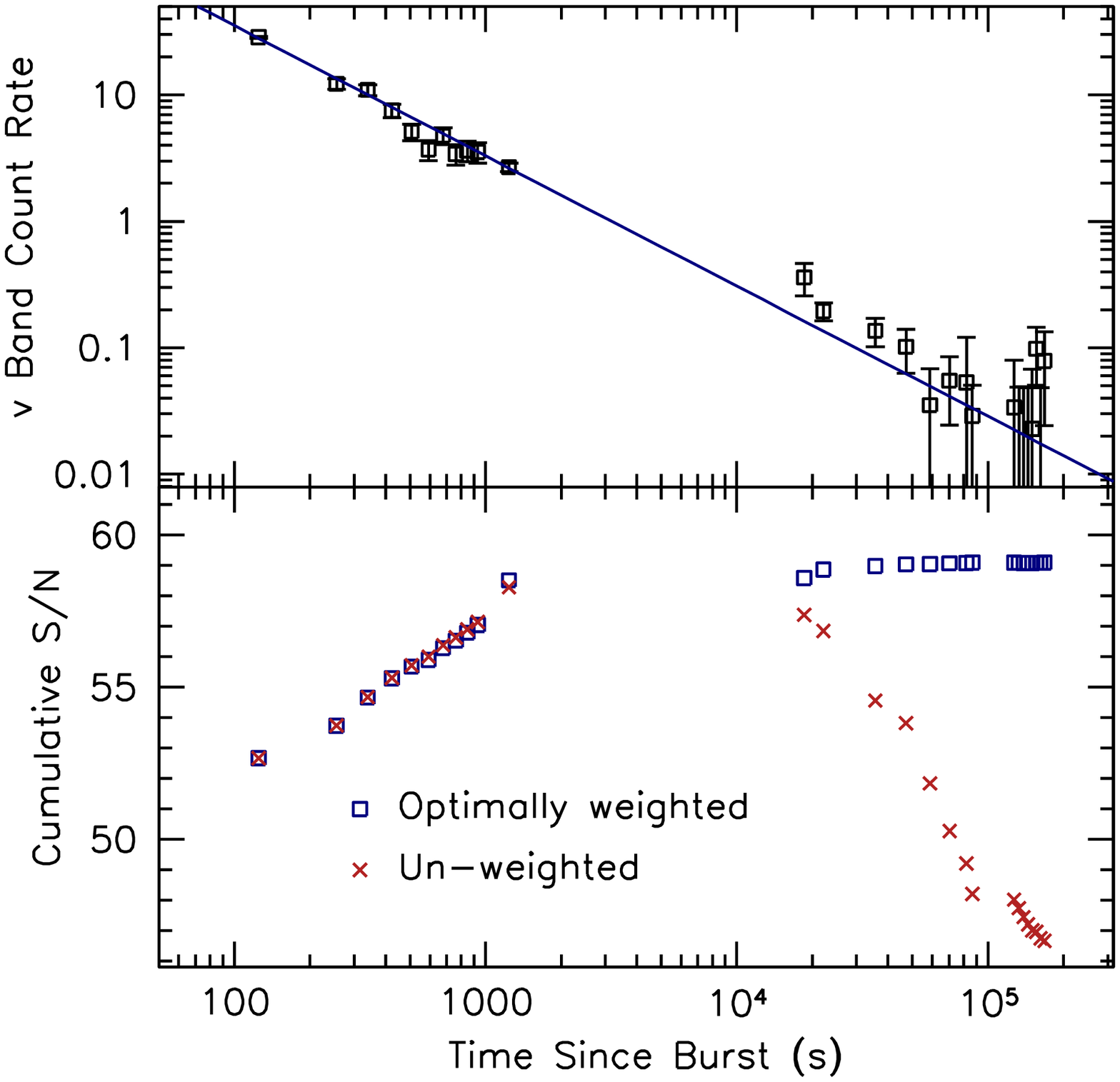}
\caption{Lightcurve (top), and cumulative $S/N$ (bottom) for UVOT
  $v$ band image data of GRB\,050525A, as a function of time since the
  burst.  Error bars on the lightcurve represent count rate uncertainties
  for measurements using individual images.  In the $S/N$ plot, red x's
  show the results of unweighted coaddition; blue squares show the results
  of optimal weighting coaddition.}
  \label{fig:050525a}
\end{figure}

The UVOT $v$ band lightcurve of GRB\,050525A is shown in the top
panel of Fig.\,\ref{fig:050525a}, and was constructed as follows.
Source plus background counts in the individual $v$ band images were
measured using a 3 arcsecond radius aperture.  A large background annulus
region, centered on the afterglow position, and excluding small regions
around any detected stars near the annulus, was used to estimate the
background counts.  Source counts were measured by subtracting the
expected average background counts inside the source extraction region
from the total counts inside that region.  A power law model was fit to
the lightcurve by finding the parameters that minimized the $\chi^{2}$
value of the fit, using the linear count rate data and the associated
uncertainties.  The best fit power law has an index $\alpha=-1.03$,
and is shown in Fig.\,\ref{fig:050525a}.  The value for the power law
index is fairly close to that found by \citet{blustin06a}, who found
a value of $\alpha = -1.14$ when a simultaneous fit was made to data
in multiple bands.  However, \citet{blustin06a} found that a single
power law does not give a full statistical description of the
optical/UV lightcurve in any band.  For this example, we have ignored
the deviations from a single power law, since they would unnecessarily
complicate the analysis, and would provide very little improvement to
the optimal coaddition of the data.

The $v$ band data were coadded without weighting, and using optimal
weighting according to equation (\ref{eq:C_totSum4}).  The GRB afterglow
counts expected in each image were calculated according to equation
(\ref{eq:csrc}), using the best fit value of the decay index in
the $v$ band.  The cumulative $S/N$ at each exposure was calculated
from equation (\ref{eq:snfinal}) for the unweighted data, and from
equation (\ref{eq:sn_wt}) for the weighted data.  The bottom panel of
Fig.\,\ref{fig:050525a} shows the cumulative $S/N$ for both the weighted
and unweighted cases.  The $S/N$ values in both cases match each other very
closely through the first $\approx 1000$s after the burst.  Within that
time range, the afterglow was still quite bright, and the noise in
the images is dominated by the counts from the afterglow itself;
optimal coaddition holds little advantage in that case.  However, at
later times, when the afterglow count rate had dropped to relatively
low values, continued unweighted coaddition of the data becomes
detrimental.  When all of the data are coadded without weighting,
the final $S/N$ is well below that of the first image by itself.
When the data are optimally coadded, the $S/N$ at later times actually
increases slightly, and levels off at a maximum value.

While coaddition of data in the case of GRB\,050525A was not necessary
for a detection, this example illustrates the danger of coadding image
data without considering the rapidly decaying nature of afterglow
lightcurves.  At best, unweighted coaddition makes no improvements,
and at worst it can severely reduce the quality of coadded data.  
Optimal coaddition nearly guarantees that the highest $S/N$ will be
returned from coadded data, regardless of how many images are coadded.
In the next two sections, we illustrate the benefits of optimal
coaddition applied to data with low $S/N$ in individual images.

\subsection{Optical Detections from Optimally Coadded Data \label{sec:newDet}}

A considerable fraction of optical GRB afterglows are not significantly
detected (at $>3\sigma$) in individual UVOT images taken with a specific
filter, but may be significantly detected in the coadded data.  An example
in which the UVOT $v$ band data reveal no afterglow in individual
images, but optimal coaddition results in a detection is GRB\,060604.
This GRB was discovered by the {\em Swift} BAT, and the spacecraft slewed
to the source position about 100s after the detection \citep{page06}.
An afterglow candidate was found in the initial UVOT {\em white} filter
image, and later confirmed in unweighted coadded UVOT {\em b} and {\em
u} images \citep{blustin06b}, and by ground-based {\em R} band images
\citep{tanvir06}.  No significant afterglow was found in any of the
individual UVOT {\em v} band images; unweighted coaddition of over 1100s
of data yielded a possible source with a significance of 2.4\,$\sigma$
\citep{blustin06b}.  This case is a good test of the optimal coaddition
technique, since we know that there was an optical afterglow in several
bands, but that the source was only marginally detected in another band.
The optimal coaddition technique was applied to the {\em v} band data
to determine if the significance of the afterglow could be improved.

\begin{figure}
\plotone{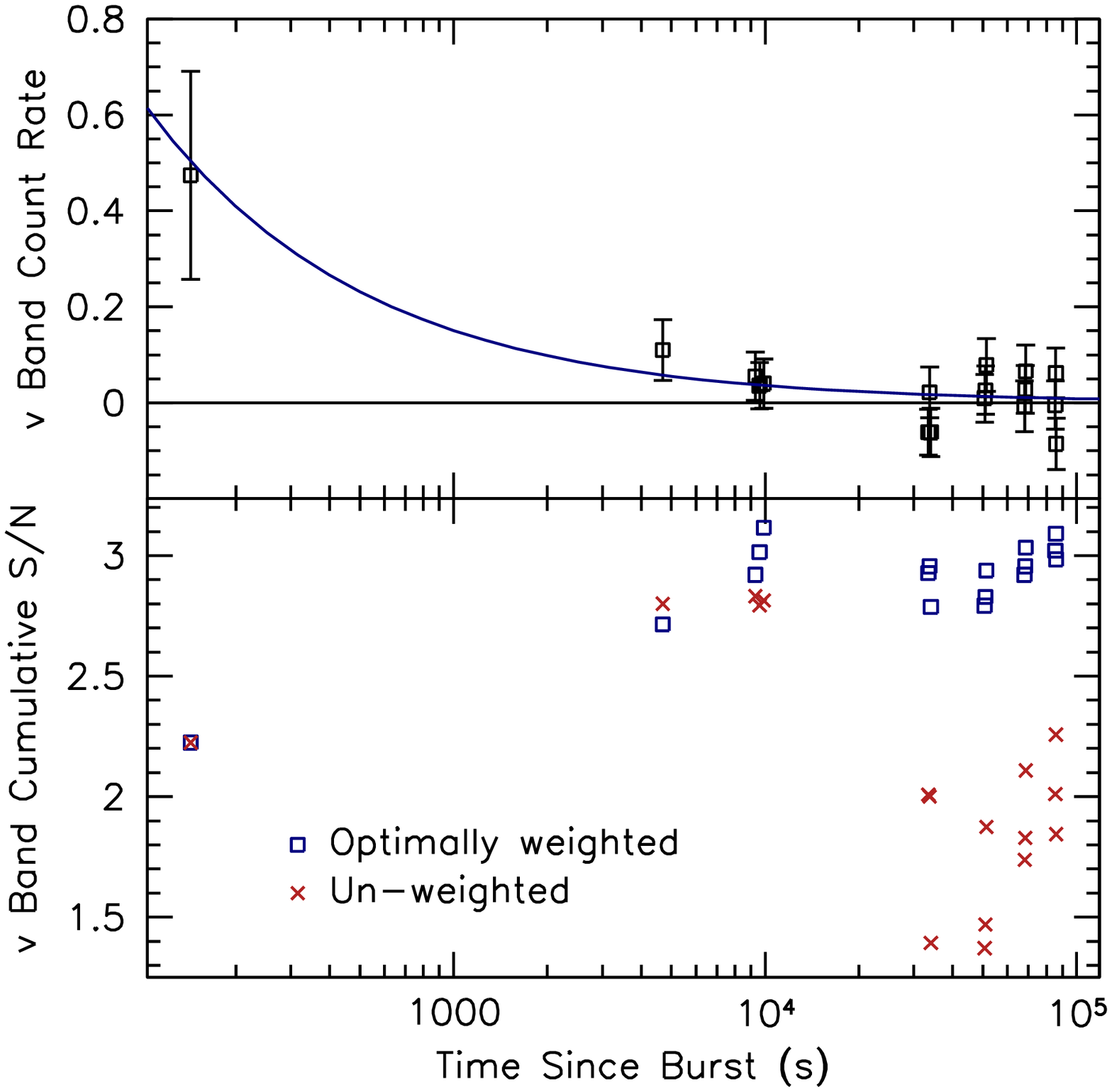}
\caption{Light curve (top), and cumulative $S/N$ (bottom) for UVOT
  $v$ band image data of GRB\,060604, as a function of time since the
  burst.  Error bars on the lightcurve represent count rate uncertainties
  for measurements using individual images.  In the $S/N$ plot, red x's
  show the results of unweighted coaddition; blue squares show the results
  of optimal weighting coaddition.}
  \label{fig:060604.1}
\end{figure}

The {\em v} band data consist of 17 images, excluding the initial very
short settling image (taken while the spacecraft is still slewing),
starting about 220\,s after the burst, and ending about $10^{5}$\,s after
the burst.  The optimal weights used in equation (\ref{eq:C_totSum1})
were determined from measurements of the source counts and background
counts in each image, and the temporal decay slope.  Source plus
background counts in the {\em v} band images were measured inside an
aperture centered on the known GRB afterglow position, with an aperture of
radius 3\,arcseconds, which is about optimal for maximizing the $S/N$ of
faint sources in UVOT images \citep{poole08,li06a}.  A large background
extraction region was selected to be near the afterglow position,
but located so that it did not contain any obviously detected stars.
Source counts were determined by subtracting the expected average
background counts inside the source extraction region from the total
counts inside that region.  The {\em v} band count rates are shown as a
function of time since burst in the top panel of Fig.\,\ref{fig:060604.1}.
The best fitting power law was determined using the technique described
at the end of \S\,\ref{sec:simulations}.  The uncertainty on the power
law slope was estimated by calculating the rms of the best slope values
in 100 simulated data sets.  In each of the simulations the counts
in the source and background regions were randomly varied according
to a Poisson distribution, with means equal to the measured values.
The best power law slope and its uncertainty are $\alpha = -0.62 \pm
0.19$; the best fitting power law is shown in Fig.\,\ref{fig:060604.1}.
The best slope value is consistent with the value found from the UVOT
$white$ band data of $\alpha = -0.62 \pm 0.32$ \citep{blustin06b},
and from the ground-based $R$ band data of $\alpha = -0.35 \pm 0.36$
\citep{tanvir06,garnavich06}.  It is clear that although none of the
individual measurements give a significant detection, the power law decay
function fit to all of the data is quite reasonable.  In this case, we
have used the fitted value of $\alpha$ because it is consistent with the
decay slopes in other bands for which there are afterglow detections,
and by design it will give the maximum weighted $S/N$.  In the absence
of a reliable fit in the $v$ band, it would have been appropriate to
adopt a value of $\alpha$ given by the results in the other bands.

\begin{figure}
\plottwo{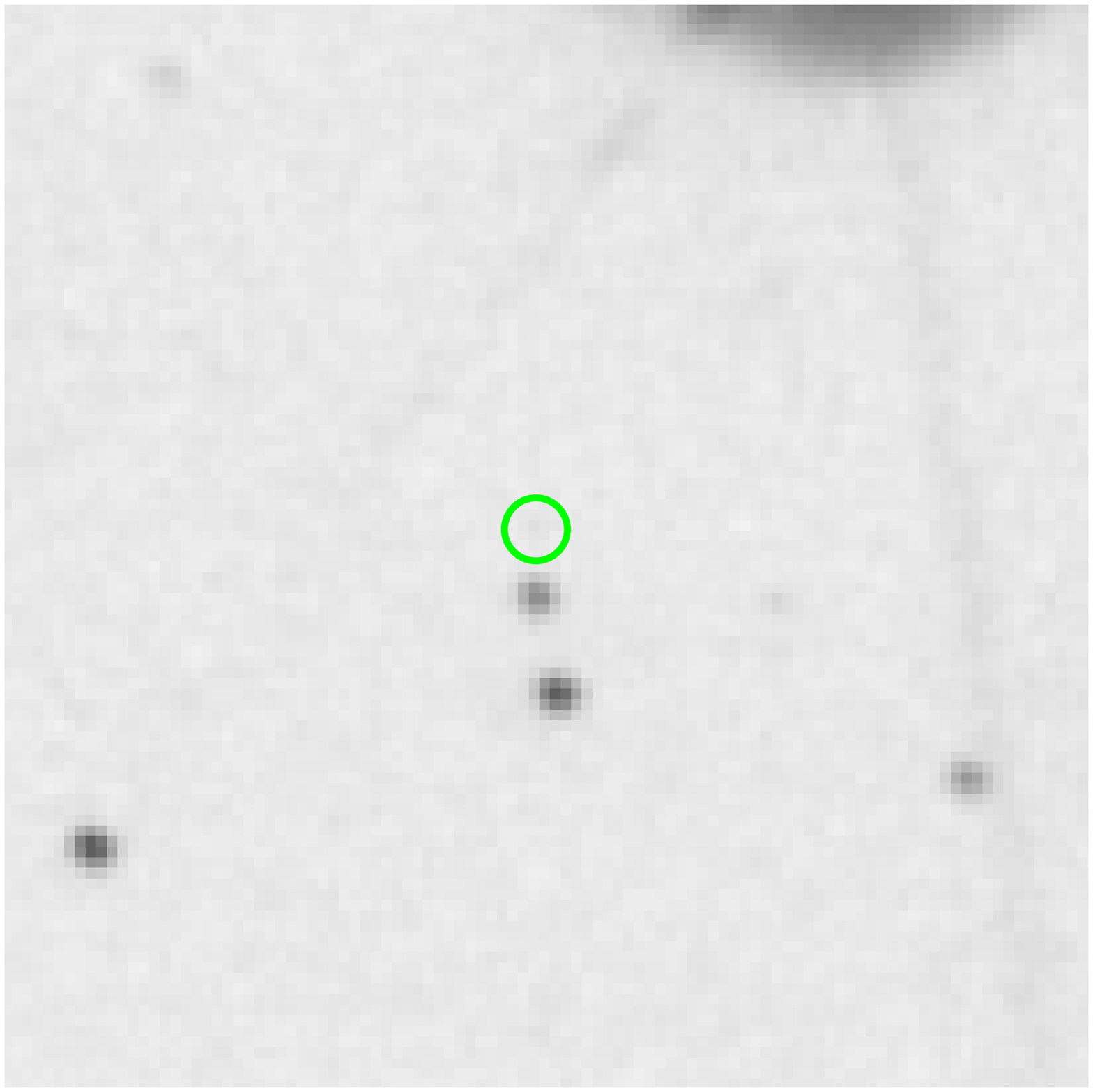}{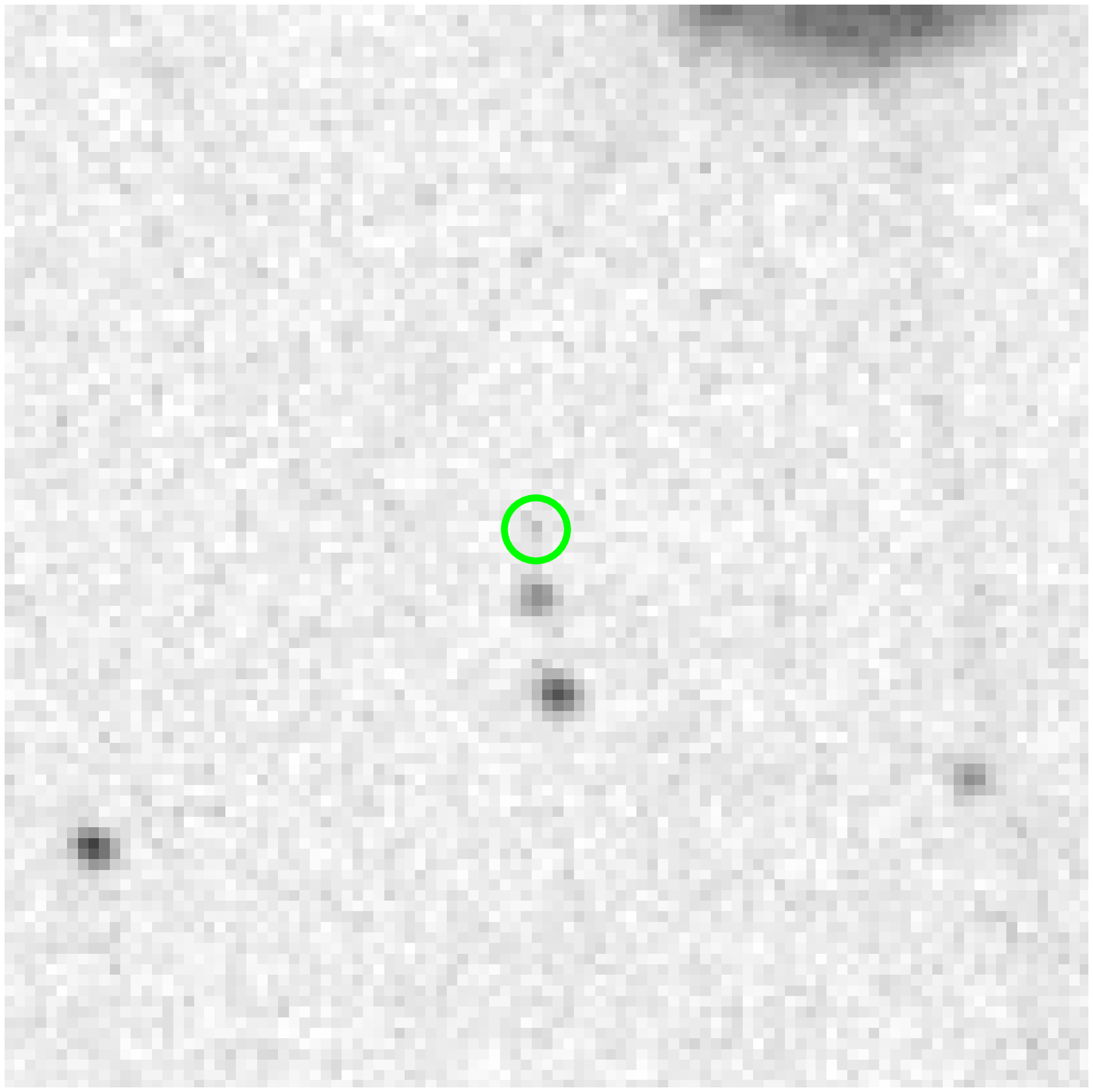}
\caption{Coadded UVOT $v$ band images of GRB\,060604. Left --- result from 
  unweighted coaddition.  Right --- result from optimally weighted
  coaddition.  Each image is $4\times4$\,arcmin on a side.  The images
  have been scaled so that the brightnesses of the field stars appear
  nearly the same in both images.  The reported
  location of the afterglow, detected in other bands, is at the center
  in each case.  The circles show the 3\,arcsec radius apertures used
  to extract the source counts.  In the unweighted case, the $S/N$ of the
  counts in the source region is about 2.0, while in the weighted case it
  is about 3.0.} \label{fig:060604.2}
\end{figure}

The cumulative $S/N$ at each exposure was calculated from equation
(\ref{eq:sn}) for the unweighted data, and from equation (\ref{eq:sn_wt})
for the weighted data.  The bottom panel of Fig.\,\ref{fig:060604.1}
shows the cumulative $S/N$ for both the weighted and unweighted cases.
The $S/N$ rises with coaddition in both cases up to the first few images.
In the unweighted case the $S/N$ drops from a high of about 2.8 to about
2.0 after coadding all of the data.  In the optimal weighting case, the
$S/N$ rises to about 3.0, and remains quite steady thereafter.  Both the
maximum unweighted $S/N$ and the final weighted $S/N$ are greater than the
value (2.4) reported by \citet{blustin06b}.  Using optimal coaddition,
the {\em v} band data by itself would have been sufficient to detect
the optical afterglow.  The individual images were coadded using no
weighting, and using the optimal weights calculated as described here.
Images centered on the afterglow position are shown for both cases in
Fig.\ref{fig:060604.2}.  The images have been scaled so that the field
stars appear to have the same brightnesses in both images; this allows a
direct comparison of the relative coadded brightness of the afterglow in
both images.  The effect of optimal coaddition is to increase the $S/N$
of the afterglow relative to unweighted coaddition.  The process also
{\em degrades} the $S/N$ of other objects in the image, since for them
the weights are not optimal.  The afterglow is difficult to see in the
unweighted coadded image, but easily seen in the weighted coadded image.

This case illustrates the potential of optimal coaddition to improve
the detection rates of GRB afterglows.  We plan to apply this method
to the entire UVOT GRB database \citep{roming08}, to determine if
new detections of other afterglows can be achieved.

\subsection{Improved Detection Limits for Optical Non-Detections
  \label{sec:dbursts}}

Optical afterglows have not been detected for many GRBs to date, including
GRBs discovered by {\em Swift} \citep{roming06}.  It is important for
detection limits to be as deep as possible in order to help understand
the ``dark burst'' phenomenon.  Also, when attempting to identify very high
redshift burst candidates, fainter detection limits in bluer bands place
tighter constraints on the range of possible redshifts.  Here we examine
the case of GRB~060923A, which is a potentially very high redshift
burst, with detections in the $K$ band, but no detections at optical
bands, including the UVOT $v$ band.

GRB~060923A was discovered by the {\em Swift} BAT, and the UVOT began
observations of the burst location about 85\,seconds after the trigger
\citep{stamatikos06}.  No afterglow was detected in any of the UVOT
images, either singly or coadded, and no afterglow was reported from
other observations in the $V$, $R$, $I$, and $J$ bands \citep{li06b,
melandri06, williams06, fox06a}.  An afterglow was found and confirmed
only in the $K$ band \citep{tanvir06, fox06a, fox06b}, suggesting
that GRB~060923A could be at a very high ($z>10$) redshift or heavily
reddened.  Subsequent very deep $R$ band observations revealed a possible
host galaxy with $R\sim25.5$ \citep{tanvir06}, which would favor the
reddening hypothesis.  Tighter constraints on the bluer band magnitudes
would provide correspondingly tighter constraints on either the redshift,
or extinction of the GRB.

As discussed in \S\,\ref{sec:detection}, optimal coaddition can be
used to find the maximum count rate (and therefore deepest magnitude)
that an afterglow could have had in the first image (where the count
rate would have presumably been the greatest) without being detected
at a given significance level.  The detection limit calculations
require the exposure start and stop times, an estimate of the afterglow
decay rate, and a measurement of the background counts at the GRB
position in each image.  UVOT images of the burst position were taken 
in the $v$ band from about 85\,s to about 11\,ks after the trigger.
Background counts in each of the images were estimated for a three
arcsecond radius aperture at the reported location of the $K$ band
afterglow.  The temporal decay rate is unknown, but reasonable
estimates can be made from several arguments.  First, the $K$ magnitude
of the afterglow dropped by at least 0.9 magnitudes from about
6.7\,ks to 86\,ks after the burst, which means the $K$ band decay
slope was less than $\alpha_{K} = -0.33$.  Second, the XRT lightcurve
faded at two distinct rates: $\alpha_{XRT1} = -2.7$ from about 80
to 400\,s, and $\alpha_{XRT2} = -1.23$ after about 4000\,s
\citep{conciatore06}.  These values bracket typical optical afterglow
decay slopes; calculations were made using all three values.

Detection limits for a $S/N$ of three in the first UVOT $v$ band
image were made for the image by itself, the unweighted coadded
image, and the optimally weighted coadded image, as described in
\S\,\ref{sec:detection}.  The count rate limits were converted to
standard UVOT $v$ band magnitudes using the zero points and aperture
corrections given by \citet{poole08}.  The results for each of the
three decay slopes are given in Table\,\ref{table:limits}.  The single
image limit is $v=20.22$.  For unweighted coaddition, there is little
($v=20.26$) or no improvement with a shallow decay slope ($\alpha =
-0.33$), and the limit actually becomes worse with steep decay slopes.
The reason for this is that noise in the later images degrades the
$S/N$ of unweighted coadded images, so that a brighter source in the
first image is required to produce a significant detection in the final
coadded image.  For a shallow slope, the optimally weighted $v$ band
limit ($v=20.42$) in the first frame is significantly deeper than the
single frame and unweighted coaddition limits.  For steep decay slopes,
the optimally weighted limit is virtually the same as the single frame
limit (and much better than the unweighted coaddition limit).  There is
little improvement over the single frame limit in this case, because
the steep lightcurve decay makes the later image weights very small,
so that they contribute very little to the final $S/N$.

The early UVOT $v$ band detection limits are the deepest reported for
GRB~060923A.  Any estimate of either a photometric redshift, or an
extinction value for this burst, must be consistent with the detection
limits.  The reported $K$ band detection measurement was made more than
$6200$\,s after the first $v$ band measurement.  Given the range of
decay slopes considered, the $v$ band magnitude limit at the time of
the $K$ band measurement would be at least 1.3 magnitudes fainter
than in the first image.  This would place even tighter constraints
on either the redshift or extinction.  The same analysis can be applied
to time series images of any non-detected GRB afterglow to improve the
detection limits.

\section{Discussion and Summary \label{sec:summary}}

Image addition for the detection of variable objects, or the improvement
of $S/N$, can benefit significantly from proper weighting.  Data from
rapidly varying sources, such as GRB afterglows, stands to benefit the
most from optimal coaddition.  In fact, it can be counterproductive {\em
not} to optimally weight the data when coadding.  We have shown how
optimal coaddition can ensure the best $S/N$ for rapidly varying GRB
afterglow data, even to the point of recovering previously undetected
sources, and can place tighter detection limits on image data sets.

We have assumed a simple power law model for the optical lightcurve; this
will clearly be inadequate in many cases \citep{oates08}.  A spectral
extrapolation from the X-ray to the optical lightcurve may provide
a reasonable estimate at early times before the forward shock peak.
The spectral index $\beta$ should vary between $\beta = (p-1)/2$
and $\beta = p/2$ where $p$ is the electron index \citep{sari98}.
Given a range of possible $p$ values, one could predict a range of
possible optical lightcurves.  However, at early times in many X-ray
lightcurves, there is a fast decay phase with superimposed flaring
activity, which rarely has a counterpart at optical wavelengths.
It is probably better in almost all cases to assume an average optical
lightcurve \citep[e.g.][]{oates08} rather than extrapolate from
an X-ray lightcurve.  In any case, it is straightforward to adapt
the coaddition technique to arbitrary lightcurves.

The technique can also be applied to observations of any variable source
(e.g.  supernovae, active galactic nuclei, cataclysmic variable stars),
as long as a reasonable estimate of the lightcurve is available.  Optimal
coaddition is most beneficial when count rates are low, background is
high, and the rate of source variability is high.  All of these factors often
apply to optical observations of GRB afterglows.  Optical observations
have been used in the examples presented here, but the technique is
equally applicable to UV, infra-red, and X-ray imaging data that has
Poisson noise characteristics.  The benefit to X-ray and possibly UV data
may not be as great as for other frequency ranges, since the background
count rates are often quite low.  However, there would likely be some
improvement in $S/N$ or detection limits, and no loss in either, when
the source count rates are comparable to the background rates.

There are other ways of improving $S/N$ that could probably be used
in combination with the optimal coaddition technique described here.
For example, we have used simple circular apertures to extract source
counts in individual images; if the image point spread function can be
determined with precision, there is the potential for optimally extracting
the counts in each image, or the final coadded image.  A related optimal
pixel weighting technique has been described by \citet{naylor98}.
In addition, if the characteristics (aside from source brightness) of
images differ significantly from frame to frame (which is usually not
the case for UVOT data), there are techniques to optimally coadd the
images to achieve improved $S/N$ or image resolution \citep{fischer94}.
As another example, the UVOT can observe in event mode, by tagging the
arrival time of individual detected photons.  This allows the possibility
of optimally weighting individual {\em events}, rather than integrated
images, which could improve the $S/N$ of afterglow data, particularly
at early post-burst times.  We also caution that the equations derived
here assume count rates high enough that Gaussian noise statistics
can be assumed, but at extremely low count rates that assumption may
not apply, and other approaches should be considered.

We will apply optimal image coaddition to all of the GRB afterglow
data obtained with the {\em Swift} UVOT.  The primary goal will be
the recovery of previously undetected afterglows, but at a minimum, the
detection limits will be improved, placing stronger constraints on
the existence of so-called dark GRBs.

\acknowledgements
Funding for the Swift program at Penn State is provided by NASA under
the contract NAS5-00136.  We acknowledge the use of public data from
the Swift data archive.

{\it Facilities:} \facility{Swift (UVOT)}


\clearpage
%
\input{tab1}

\clearpage

%
%
%
%
%
%
%

\end{document}

%% file: tab1.tex
\begin{deluxetable}{lccc}

\tablewidth{0pt}

\tablecaption{GRB~060923A UVOT $v$ band magnitude limits for the first image.}

\tablehead{
  \colhead{$\alpha$} &
  \colhead{Single} &
  \colhead{Unweighted} &
  \colhead{Optimal} \\
  \colhead{} &
  \colhead{Image} &
  \colhead{Coaddition} &
  \colhead{Coaddition}
}

\startdata
  -0.33 & 20.22 & 20.26 & 20.42 \\
  -1.23 & 20.22 & 19.60 & 20.22 \\
  -2.70 & 20.22 & 19.53 & 20.22
\enddata
\label{table:limits}
\end{deluxetable}